\documentclass[a4paper, 10pt, oneside, onecolumn,preprint]{elsarticle}

\usepackage{graphicx}
\usepackage{amssymb}
\usepackage{lineno}
\usepackage{booktabs}

\usepackage{alltt}
\usepackage{ifthen}
\usepackage{float}
\usepackage{calc}
\usepackage{stmaryrd,amsmath, amsthm, amssymb}
\usepackage{listings}

\newcommand{\foops}[1]{\mbox{\texttt{#1}}}
\newcommand{\cc}[1]{\ensuremath{#1}}

\newcommand{\bfoo}{\bfooc{true}}
\newcommand{\efoo}{\efooc{true}}
\newcommand{\bfooc}[1]
    {\ifthenelse{\equal{#1}{true}}{\begin{alltt}}{\begin{forget}}}
\newcommand{\efooc}[1]
    {\ifthenelse{\equal{#1}{true}}{\end{alltt}}{\end{forget}}}

\DeclareSymbolFont{mytypewriter}{OT1}{cmtt}{m}{n}
\DeclareMathSymbol{\lbracett}{\mathopen}{mytypewriter}{'173}
\DeclareMathSymbol{\rbracett}{\mathclose}{mytypewriter}{'175}

\newcommand{\code}[1]{\foops{#1}}
\newcommand{\br}{\ensuremath{\rbracett}}
\newcommand{\bl}{\ensuremath{\lbracett}}
\newcommand{\fact}{\ensuremath{\mathbf{fact}}}
\newcommand{\sig}{\ensuremath{\mathbf{sig}}}
\newcommand{\set}{\ensuremath{\mathbf{set}}}

\newcommand{\one}{\ensuremath{\mathbf{one}}}
\newcommand{\no}{\ensuremath{\mathbf{no}}}

\newcommand{\inAlloy}{\ensuremath{\mathbf{in}}}
\newcommand{\extendsAlloy}{\ensuremath{\mathbf{extends}}}

\newcommand{\alll}{\ensuremath{\mathbf{all}}}

\newcommand{\some}{\ensuremath{\mathbf{some}}}

\journal{Science of Computer Programming}

\begin{document}
\nolinenumbers
\begin{frontmatter}


\title{A Domain-Specific Language for Verifying Software Requirement Constraints}



\author{Marzina Vidal}
\ead{marzina@ufcg.edu.br}
\address{Federal University of Campina Grande, Brazil}

\author{Tiago Massoni}
\ead{massoni@computacao.ufcg.edu.br}
\address{Federal University of Campina Grande, Brazil}

\author{Franklin Ramalho}
\ead{franklin@computacao.ufcg.edu.br}
\address{Federal University of Campina Grande, Brazil}

\begin{abstract}
Software requirement analysis can certainly benefit from prevention and early detection of failures, in particular by some kind of automatic analysis. 
Formal methods offer means to represent and analyze requirements with rigorous tools, avoiding ambiguities and allowing automatic verification of requirement consistency. 
However, formalisms often clash in the culture or lack of skills of software analysts, making them challenging to apply.
In this article, we propose a Domain-Specific Language (DSL) based on Set Theory for requirement analysts. The Graphical InvaRiant Language (GIRL) can be used to specify software requirement structural invariants, with entities and their relationships. 
Those invariants can then have their consistency evaluated by the Alloy Analyzer, based on a \emph{mapping semantics} we provide for transforming GIRL models into Alloy specifications with no user intervention.
With a prototypical language editor and transformations implemented into an Eclipse plugin, we carried out a qualitative study with requirement analysts working for a government software company in Brazil, to evaluate usability and effectiveness of the GIRL-based analysis of real software requirements.
The participants were able to effectively use the underlying formal analysis, since 79 out of 80 assigned invariants were correctly modeled.
While participants perceived as low the complexity of learning and using GIRL's simplest, set-based structures and relationships, the most complex logical structures, such as quantification and implication, were challenging. 
Furthermore, almost all post-study evaluations from the participants were positive, especially as a tool for discovering requirement inconsistencies.
\end{abstract}


\end{frontmatter}


\section{Introduction}
\label{intro}

Accurate and consistent requirements produce quality software at an appropriate cost. 
Many software problems arise from failures in the way people obtain knowledge, document, agree, and modify product requirements~\cite{Wiegers-book}. 
Incomplete requirements, followed by communication failures between the project team and the client, and change of targets, are currently the three most frequent problems in requirements engineering~\cite{naming-pain}.
To minimize these problems, we need to improve \emph{prevention and early detection of failures}; it is cheaper to fix a concept than to fix the code under test, or worse, to correct the code of running software.

When seeking for failures in requirements, one could use \emph{formal methods} to represent and analyze requirements with rigorous tools, avoiding ambiguities and allowing automatic verification of requirement consistency. However, the means to apply such formalisms often clash in the culture or lack of skills of software analysts~\cite{Wiegers-book}, making those methods challenging to apply. Some proponents of formalisms indeed consider the need for greater convergence or cooperation between the cultures of formal methods and intuition~\cite{hoare-collusion}, such as using UML-like or Domain-Specific Languages (DSL) with a reliable and hidden formal infrastructure~\cite{mythbusters}.

In addition, one could consider the superiority of graphic over text, as explained by the different ways the mind processes information \cite{physics}. 
Therefore, the adoption of a well-designed visual notation promotes learning, communicability, and precision. 
In this perspective, we notice that the visual notation of \emph{Set Theory} in mathematics is a simple option that favors the semantic transparency~\cite{physics}, since it is widely known and produces a rapid inference of its meaning.

Following this conciliation between formalism and intuition, we propose a DSL based on Set Theory for requirement analysts. 
The Graphical InvaRiant Language (GIRL) can be used to specify software \emph{structural invariants}, with entities and their relationships. 
Those invariants can then have their consistency evaluated by the Alloy Analyzer~\cite{alloy-book}, based on a \emph{mapping semantics} we provide by automatically transforming GIRL models into Alloy specifications.
If the GIRL model is consistent, instances of entities and relationships from the original model are presented graphically. 

With a prototypical language editor and transformations implemented into an Eclipse plugin, we carried out an empirical study, with requirement analysts working for a government software company in Brazil, to evaluate \emph{usability and effectiveness} of the GIRL-based analysis of real software requirements. Using as strategy \emph{judgment tasks}~\cite{StolFitzgerald}, ten (10) participants assessed language understanding, verification effectiveness, and general usefulness, after receiving appropriate training.
The data provides positive and negative evidence on the perception about GIRL, which we could speculate as recurrent in graphical notations with hidden formal verification.

While participants perceived as low the complexity of learning and using its simplest, set-based structures and relationships, the most complex logical structures, such as quantification and implication, were considered as challenging. However, almost all invariants (eight invariants per each participant (80 invariants in total) -- only one was incorrectly delivered) were correctly modeled, as all participants were able to effectively use the underlying formal analysis provided by the Alloy Analyzer.  
Also, almost all post-study evaluations from the participants were positive, especially as a tool for discovering frequent requirement inconsistencies.

The relevance of this study is three-fold: first, in economic terms, GIRL allows the elaboration of verifiable models, potentially minimizing development costs. 
Second, we see potential in circumventing the mythical resistance of software developers to the use of mathematical notation, even though their training curricula contain topics in mathematics and formal methods~\cite{mythbusters}. 
Finally, the involvement of software professionals modeling real requirements is undoubtedly essential for assessing software verification costs and benefits.

This article is organized as follows. Section~\ref{back} refers to a brief background information on Alloy and DSLs, while Section~\ref{girl} presents the GIRL language. In Section~\ref{map}, we introduce the translation semantics in Alloy, whereas in Section~\ref{tool} we give an overview about the provided tool apparatus. In Sections~\ref{eval} and ~\ref{results}, we report and discuss the empirical study we carried out for evaluating GIRL. Section~\ref{related} compares GIRL with previous research approaches, while we discuss the implications of this work and its future in Section~\ref{conclusions}.

\section{Background}
\label{back}

In this section, we provide a brief overview about the concepts required for the proposal.

\subsection{Alloy}

\emph{Alloy} \cite{alloy-book} is a formal object-oriented modeling language, based on first-order logic and a notation called \emph{relational calculus}, that gives a mathematical notation for specifying objects and their relationships. Alloy models are similar to UML class diagrams combined to Object Constraint Language (OCL)~\cite{ocl}, but Alloy has a simpler syntax, type system and semantics, being designed for automatic analysis, which motivated us to choose the language for this research work. 
The language assumes a universe of elements partitioned into subsets, each of which associated with a defining type. An Alloy model contains a sequence of \textit{paragraphs}; one kind of paragraph is called a \emph{signature}, which is used for defining a new type. These instances can be related by \emph{relations} declared in the signatures. A signature paragraph introduces a basic type and a collection of relations, along with their types and other constraints on the values that they relate.

For a simple file system object model, the following Alloy fragment defines signatures for \code{FSObject} and \code{Name}. The keyword \code{\sig} declares a signature with a name. Signature \code{Name} is an empty signature, while \code{FSObject} declares two relations. For example, every instance of \code{FSObject} is related to exactly one instance of signature \code{Name} by the relation \code{name} -- the keyword \code{\one} denotes a total function. Also, every file system object may have \code{contents}; it is optional, and maybe contains more than one instance, since it is annotated with the \code{\set} keyword, which establishes no constraints on the relation.
\bfoo{\small
  \cc{\sig} Name \bl\br
  \cc{\sig} FSObject \bl
    name: \cc{\one} Name,
    contents: \cc{\set} FSObject
  \br
}\efoo%

In Alloy, one signature can extend another -- with the \code{\extendsAlloy} keyword --, establishing that the extended signature is a subset of its supersignature. Signature extension introduces a subtype, establishing that each subsignature is disjoint. The following fragment shows signatures called \code{File} and \code{Dir}, that are disjoint (\code{FSObject} instances may be exclusively files or directories). In addition, \code{Root} is a subtype of \code{Directory}.
\bfoo{\small
  \cc{\sig} File \cc{\extendsAlloy} FSObject \bl\br
  \cc{\sig} Dir \cc{\extendsAlloy} FSObject \bl\br
  \cc{\sig} Root \cc{\extendsAlloy} Dir \bl\br
}\efoo%

This specification can be further constrained with invariants, for complex domain rules concerning the declared signatures and relations. 
For this purpose, it can be enriched with formula paragraphs called \textit{fact}, which is used to package formulae that always hold for the model. 
The following fragment introduces a fact establishing general properties about the file system. 
The first formula states, using the \code{\one} keyword, that there must be exactly one \code{Root} instance in every file system. 
Next, the second formula defines that from all file system objects, only directories may present contents. In the expression \code{(FSObject-Dir).contents}, the join operator (‘.’) represents relational dereference (in this case, yielding contents from the set of instances resulting from \code{FSObject} instances that are not directories, with symbol \code{-} as set difference). The \code{no} keyword establishes that the expression that follows results in an empty set, which gives the constraint the following meaning: only directories have contents in the file system.
\bfoo{\small
  \cc{\fact} \bl
    \cc{\one} Root
    \cc{\no} (FSObject-Dir).contents
  \br
}\efoo%


Alloy was simultaneously designed with a fully automatic tool that can simulate models and check properties about them -- the Alloy Analyzer~\cite{alloy-book}. The tool translates the model to be analyzed into a boolean formula, and this formula is solved using SAT solvers. The analysis consists in binding instances to signatures and relations, searching for a combination of values that make the translated boolean formula true.

One of the analysis, namely \textit{simulation}, generates structures without requiring the user to provide sample inputs or test cases. If the tool finds a configuration of instances making the formula true, a valid \emph{interpretation} for the model is determined. In our example, we can use the \code{abstract} predicate in order to simulate the model in the tool, with the \code{run} command. The complement \code{for 3} constrains the simulation to work on a scope of at most three instances for each signature; the analysis is limited to a number of instances, being sound and complete up to that specific number~\cite{alloy-book}.
\bfoo{\small
  run abstract for 3
}\efoo%

\begin{figure}[ht]
\begin{center}
\leavevmode
\scalebox{0.8}{
\includegraphics{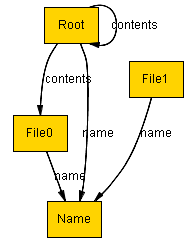}}
\caption{An interpretation for the file system model}
\label{fig:instance1}
\end{center}
\end{figure}

In this example the model is consistent, as at least one interpretation is found, as showed in Figure~\ref{fig:instance1}, excerpted from the actual tool output. Each box represents an instance with the name of the corresponding signature; for more than one instance, numbers are concatenated to the signature name. A major benefit of this analysis is that modeling errors, such as missing constraints, can be easier to find by visualizing possible interpretations. In this case, two design decisions commonly seen in file systems are unspecified: a directory having itself as contents (\code{Root}) and orphan files lacking a parent directory (\code{File1}).

These two problems can be tackled by adding two more invariants as a fact to the model. The first states that no directory can be related to itself by \code{contents}, \textit{even indirectly}; the symbol \code{\^} yields the \textit{transitive closure} of the relation. Additionally, the second invariant defines that every file is contained within a directory; keywords \code{\some} and \code{\inAlloy} represent existential quantification and element inclusion in a set, respectively. While \code{\alll} represents the universal quantifier, \code{\no} represents its negation.
\bfoo{\small
	\cc{\fact} \bl
	  \cc{\no} d:Dir | d \cc{\inAlloy} d.^contents
	  \cc{\alll} f:File | \cc{\some} d:Dir | f \cc{\inAlloy} d.contents
	\br
}\efoo%

\begin{figure}[ht]
\begin{center}
\leavevmode
\scalebox{0.8}{
\includegraphics{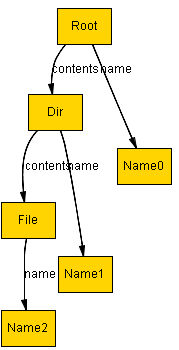}}
\caption{Interpretation for the modified model}
\label{fig:instance2}
\end{center}
\end{figure}

If additional executions are performed, one of the possible results is depicted in Figure~\ref{fig:instance2}, confirming the effects of the introduced invariants.

\subsection{DSL and MDD Transformations}
Domain-Specific Languages (DSLs) are specialized languages built to cover a specific domain. They express abstractions to a domain by means of textual or graphical notations, tables and symbols~\cite{DslEngineering2013}.
For instance, SQL is a DSL specific to create and manipulate relational schemas on databases. SQL has constructors specific to that domain, such as \textit{create table}, \textit{insert into}, \textit{select where}, that are very expressive to create, populate and query tables, respectively.

A DSL can be classified as an internal or an external DSL. The former relies on a compiler already provided to an existing language, such as JUnit (a DSL for unit testing)~\cite{Tahchiev:2010:JAS:1893022} or Ruby on Rails~\cite{DslEngineering2013}, both relying on Java compiler. On the other hand, the latter relies on a compiler fully built to the DSL, independently of the target language to which it is compiled. SQL, CSS and HTML~\cite{DamyanovSukalinska2015} are examples of external DSLs.

DSLs favor validation and verification activities because they capture and express the essential of a domain, abstracting irrelevant details. Therefore, some DSLs are built specifically to allow complex formal analysis~\cite{DslEngineering2013}.
In Model-Driven Development (MDD)~\cite{Stahl:2006:MSD:1196766}, the abstract syntax is provided by a meta-model that may also contain the static semantics. For instance, usually a MOF~\cite{omg2013mof} meta-model is a grammar that is accompanied by OCL constraints that specify well-formedness rules on the language constructors. 
Additionally, MDD transformations provide the mapping responsible for compiling the DSL to a target language. These transformations are classified as a model-to-model (M2M) transformation or as a model-to-text (M2T) transformation. The former maps a source model to a target model, both usually represented as an XMI~\cite{XMI:2015} format. In case of the target language having a textual representation, the latter is further applied to generate a compilable target code.

In this work, we provide GIRL as an external DSL for visually specifying requirement invariants. An abstract syntax is provided by the MOF GIRL meta-model also proposed in this work. In order to allow indicating whether the GIRL model is satisfiable or not, we have mapped GIRL to Alloy by means of MDD transformations (both M2M and M2T). We have adapted the available Alloy meta-model~\cite{AlloyMetamodel:2019}. 

We provide a set of QVT (Query/View/Transform)~\cite{omg2011qvt} M2M transformations that specify how the source model (GIRL) is translated to the target model (Alloy). The resulting Alloy model is in fact a XMI representation that is further transformed into Alloy textual specification by means of Acceleo~\cite{acceleo} M2T transformations. 


\section{The GIRL Language}
\label{girl}

GIRL is an external DSL with a visual syntax whose semantics is provided by means of a mapping to Alloy (domain semantics) introduced in the next section. The GIRL syntax underlies on set theory, especially Venn diagrams, to represent sets and their operations~\cite{set-theory}.
GIRL combines first order logic relations, relational calculus and object-oriented concepts, such as sets, relations, quantifiers (universal and existential), logical operators (implies, and, or), transitive closure, etc.

The core concept in GIRL is an invariant. It comprises the constraints to be specified on the model. The constraints in turn can be classified as primary or composed structures. The former allows specifying (i) An Entity to which constraints are stated; or (ii) An Integer primitive type. The latter comprises elements that are applied on GIRL elements (primary or composed), such as containment relation, set operation, relational operation, quantifiers, implications and cardinalities. 

In the sequel, we introduce the abstract and concrete syntax provided to GIRL. 

\subsection{Abstract Syntax}

\begin{figure}[]
    \centering
    \includegraphics[width=1\textwidth]{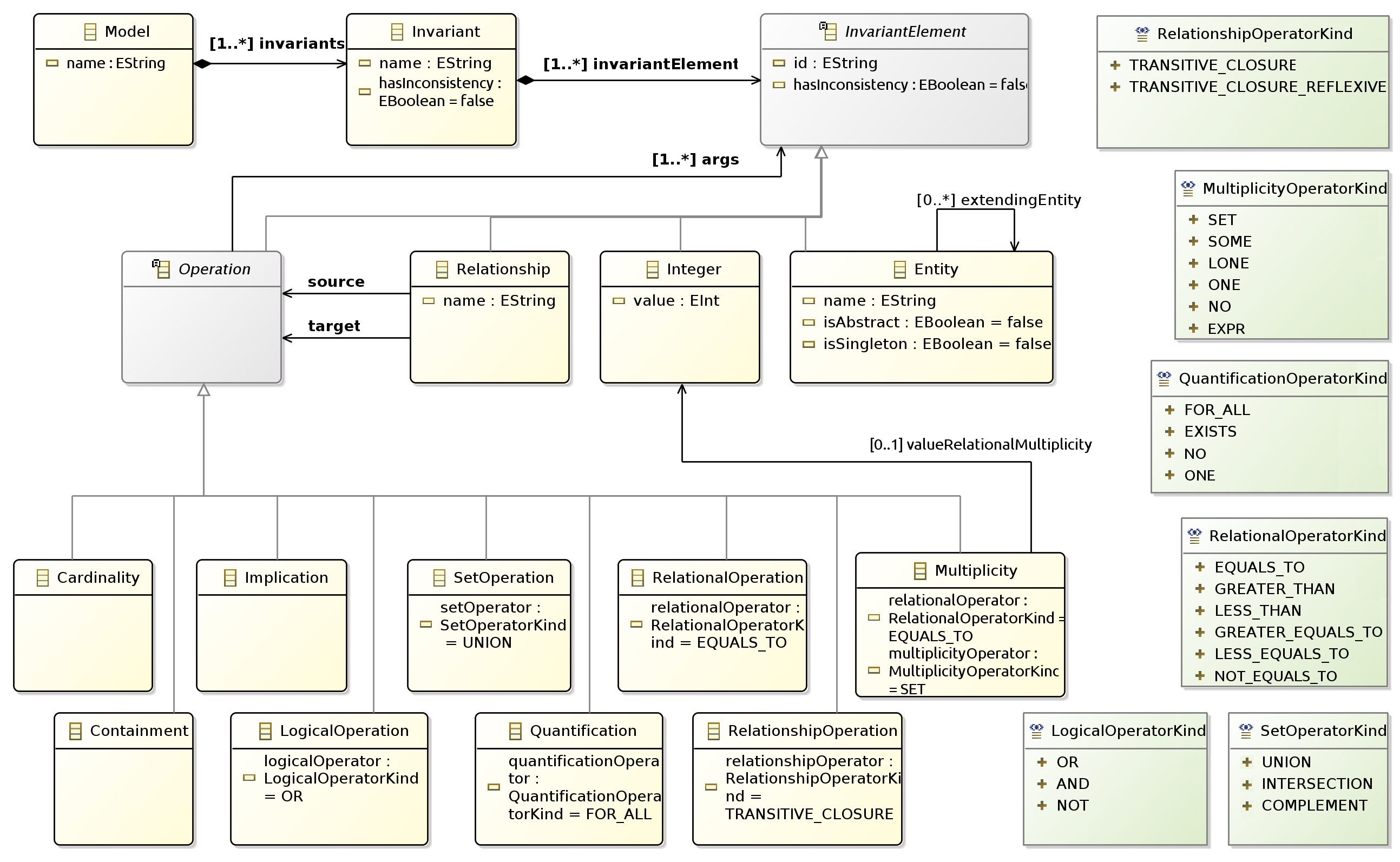}
    \caption{The MOF GIRL meta-model}
    \label{fig:GIRL_Metamodel}
\end{figure}

The GIRL abstract syntax is provided by the MOF meta-model shown in Figure ~\ref{fig:GIRL_Metamodel}.  As one can note, a GIRL model is composed of invariants, that, in turn, are composed of elements that can be classified as an \textit{Entity}, an \textit{Integer}, a \textit{Relationship} or an \textit{Operation}. The two latter are composed of other elements. An \textit{Operation} requires one or more arguments, whereas an \textit{Relationship} requires exactly one \textit{source} and one \textit{target} operation.

The GIRL operations are applied on elements according to the \textit{Operation} type that can be: 

\begin{itemize}
    \item \textit{Cardinality} - indicates the amount of elements an entity has;
    \item \textit{Containment} - states that an element is contained into another one;
    \item \textit{Implication} - covers the implies operator to be applied on two elements (premise and conclusion);
    \item \textit{LogicalOperation} - covers the logical operators (and, or, not) to be applied on one or more elements;
    \item \textit{Quantification} - covers four kinds of quantifiers operators (existential, universal, no and one);
    \item \textit{RelationalOperation} - covers the relational operators ($>$, $<$, $\geq$, $\leq$, =);
    \item \textit{Multiplicity} - defines the cardinality of an entity, a set operation, a relationship or a quantification operation;
    \item \textit{SetOperation} - covers the set operators (union, intersection or complement);
    \item \textit{RelationshipOperation} - covers the application of relations where a resulting relationship is derived from a source relationship.
\end{itemize}

\textit{Containment}, \textit{LogicalOperation}, \textit{Implication}, \textit{Quantification} and \textit{RelationalOperation} are boolean operations, whereas \textit{Cardinality} is an integer operation. Additionally, there exists the \textit{Entity} type associated to an \textit{Entity}, a \textit{SetOperation} or a \textit{RelationshipOperation}. Finally, a \textit{Relationship} or a \textit{RelationshipOperation} may result in a Relationship type.

 \subsection{Concrete Syntax}

The GIRL graphical syntax is presented in the following by means of illustrations on a bank account domain example. It is important to emphasize that some graphical element features define its semantics, such as its shape, color, position, texture, brightness and direction~\cite{physics}. In addition, in order to respect the semantic transparency principle, some visual representations are chosen to better reflect its meaning~\cite{physics}. 

\begin{figure}[htbp]
    \centering
    \includegraphics[width=0.3\textwidth]{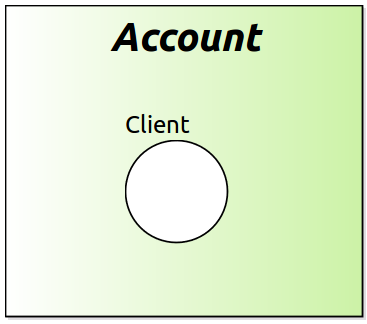}
    \caption{GIRL invariant and entity}
    \label{fig:GIRL_Invariant&Entity}
\end{figure}

\textbf{Invariant} - A GIRL invariant contains the constraint to be applied on one or more element(s). Every invariant has a context. Its notation is a quadrilateral identified with a name (the context) and filled with a gradient color from white to green. Figure~\ref{fig:GIRL_Invariant&Entity} states an invariant applied to the \textit{Account} context.

\textbf{Entity} - An Entity is a structure that may be a set or a class, both having zero or more elements. Its notation is a white circle with solid black border, similarly to the sets in Venn diagrams~\cite{set-theory}. Figure~\ref{fig:GIRL_Invariant&Entity} shows a \textit{Client} as an Entity. 

It is important to emphasize that whenever an entity occurs — in one of the following contexts: relationship, implication, quantification, or in set, containment, logical or relational operations —  it must also appear on the left side into the invariant, as showed in the following illustrations.

\textbf{Abstract and Singleton Entities} - An abstract entity is a container entity. An entity (abstract or not) may be extended by one or more entities that must be mutually disjoint. Therefore, the extending entities are subsets of the extended entity.

\begin{figure}[]
    \centering
    \fbox{\includegraphics[width=0.9\textwidth]{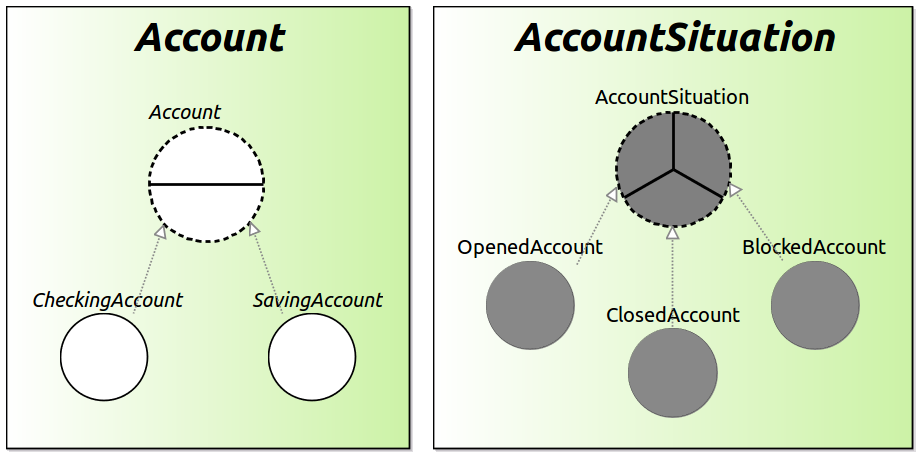}}
    \caption{GIRL abstract and singleton entities}
    \label{fig:GIRL_Abstract&SingletonEntity}
\end{figure}

The cardinality of an abstract entity is the number of child entities that extend it. Its notation is a white circle with dashed black border. The left side of Figure~\ref{fig:GIRL_Abstract&SingletonEntity} shows two children entities (\textit{CheckingAccount} and \textit{SavingAccount}) that extend the abstract entity named \textit{Account}.

On the other hand, a singleton entity must have exactly one element. Its notation is a gray circle, as the entities \textit{BlockedAccount}, \textit{OpenAccount} or \textit{ClosedAccount} shown on the right side in Figure~\ref{fig:GIRL_Abstract&SingletonEntity}. Observe that an entity can not simultaneously be abstract and a singleton.

\begin{figure}[]
    \centering
    \includegraphics[width=0.6\textwidth]{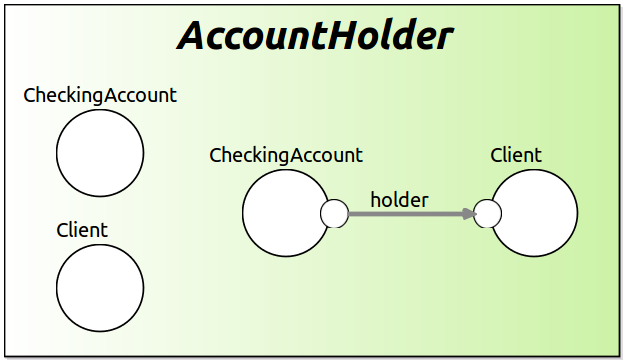}
    \caption{GIRL relationship}
    \label{fig:GIRL_Relationship}
\end{figure}

\begin{figure}[]
    \centering
    \fbox{\includegraphics[width=0.8\textwidth]{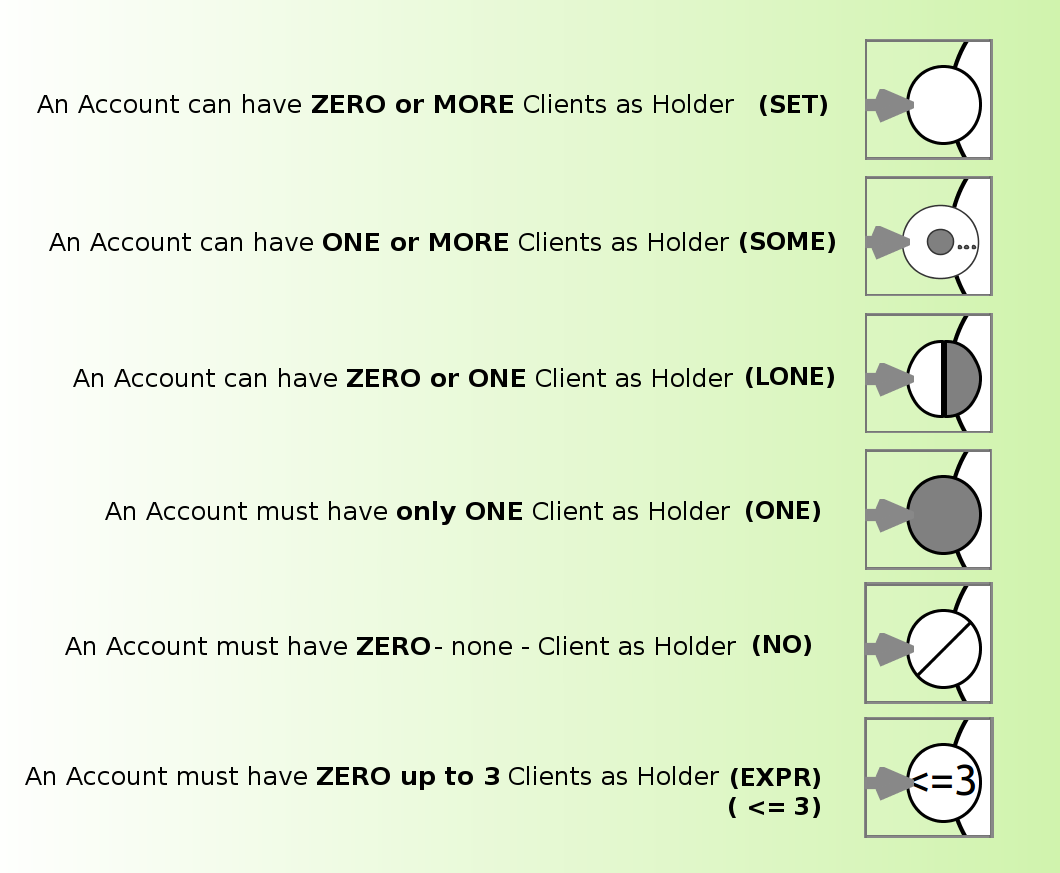}}
    \caption{GIRL relationship multiplicities}
    \label{fig:GIRL_RelationshipMultiplicities}
\end{figure}

\textbf{Relationship} - An entity can be related to another one by means of a named relationship. In particular, a GIRL relationship involves only two member elements, where each one owns a multiplicity in the relationship. These elements can be an entity, a set operation, a quantification or a relationship operation. It can still be reflexive, \textit{i.e.} the \textit{source} and the \textit{target} elements are the same. Its notation is a gray directed arrow from the source element (left side) to the target element (right side). Figure~\ref{fig:GIRL_Relationship} shows the relationship named \textit{holder} from the \textit{CheckingAccount} entity to the \textit{Client} entity. You can note that there exists small circles positioned between the element and the arrow in both sides. They indicate the multiplicity played by that element in the relationship. The multiplicity values that can occur in relationships are illustrated in Figure~\ref{fig:GIRL_RelationshipMultiplicities}. 

\begin{figure}[]
    \centering
    \includegraphics[width=0.45\textwidth]{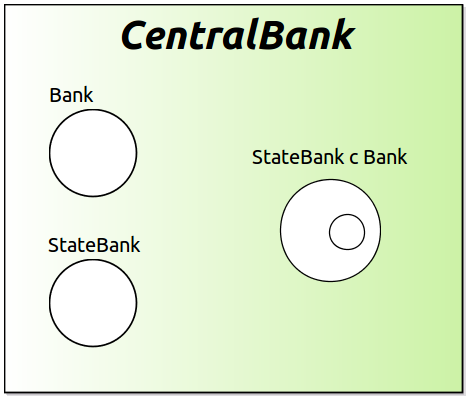}
    \caption{GIRL containment operation}
    \label{fig:GIRL_ContainmentOperation}
\end{figure}

\textbf{Containment} - The containment operation allows specifying that an entity (atomic or composite) is contained into another one. Its notation is a smaller circle inside a larger circle both filled with white color and with black borders. Additionally, the `c' operator is applied to the subset and the superset. For instance, Figure~\ref{fig:GIRL_ContainmentOperation} shows that \textit{CentralBank} is contained into a \textit{Bank}.

\begin{figure}[]
    \centering
    \includegraphics[width=0.6\textwidth]{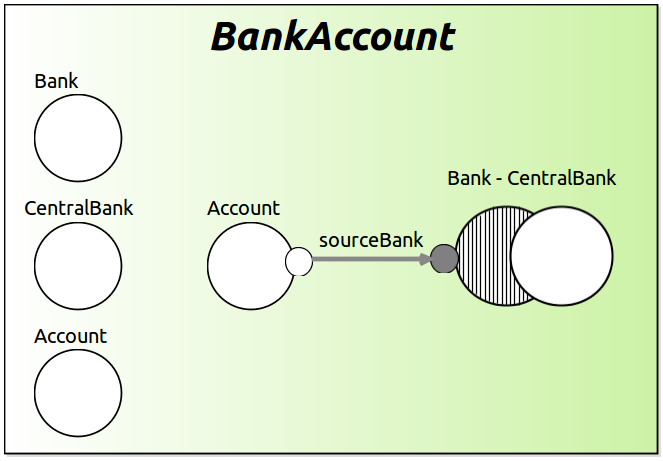}
    \caption{GIRL set operation}
    \label{fig:GIRL_SetOperation}
\end{figure}

\textbf{Set Operation} - GIRL provides the following set operations: \textit{Union, Intersection} and \textit{Complement}. The syntax is the same adopted by Venn diagrams. Figure~\ref{fig:GIRL_SetOperation} shows 
the \textit{CentralBank}, which is a singleton
Additionaly, there is an invariant (right side) that does not allow that an \textit{Account} has a relationship with the {CentralBank}. This constraint is provided by means of the \textit{complement} set operation.

\begin{figure}[]
    \centering
    \includegraphics[width=0.6\textwidth]{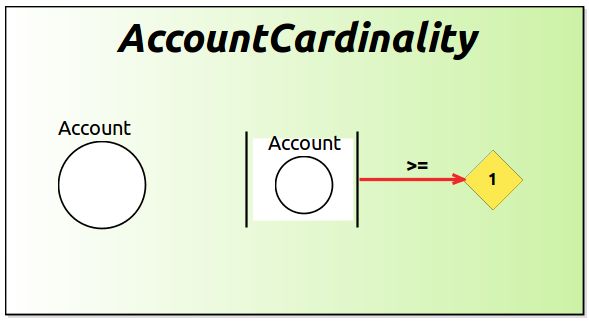}
    \caption{GIRL cardinality}
    \label{fig:GIRL_Cardinality}
\end{figure}

\textbf{Cardinality} - The cardinality of an entity specifies its size, \textit{i.e.} the amount of instances it has. Its notation is the same used in set theory, as shown in Figure~\ref{fig:GIRL_Cardinality}, where it is stated that an \textit{Account} must have one or more instances.

\textbf{Relational Operation} - A relational operation allows comparing two entities by means of a relational operator. The notation is a red arrow identified by the relational operator, as shown in Figure~\ref{fig:GIRL_Cardinality}.
 
 \textbf{Integer} - GIRL provides only the Integer primitive type. Its notation is a yellow diamond (with black borders) containing the integer value, as shown in Figure~\ref{fig:GIRL_Cardinality}.

\textbf{Logical Operation} - GIRL provides the following logical operators: conjunction, disjunction and negation. They are applied on GIRL boolean expressions. Its notation is a white rectangle as follows:
\begin{itemize}
    \item AND - with black borders identified with the term AND in the orange header. It may contain two or more elements separated by a solid line in black color;
    \item OR - Similar to the AND operator, but identified with the term OR in the blue header;
    \item NOT - Rectangle with red edges, identified with the term NOT in the header.
\end{itemize}

\begin{figure}[]
    \centering
    \includegraphics[width=1\textwidth]{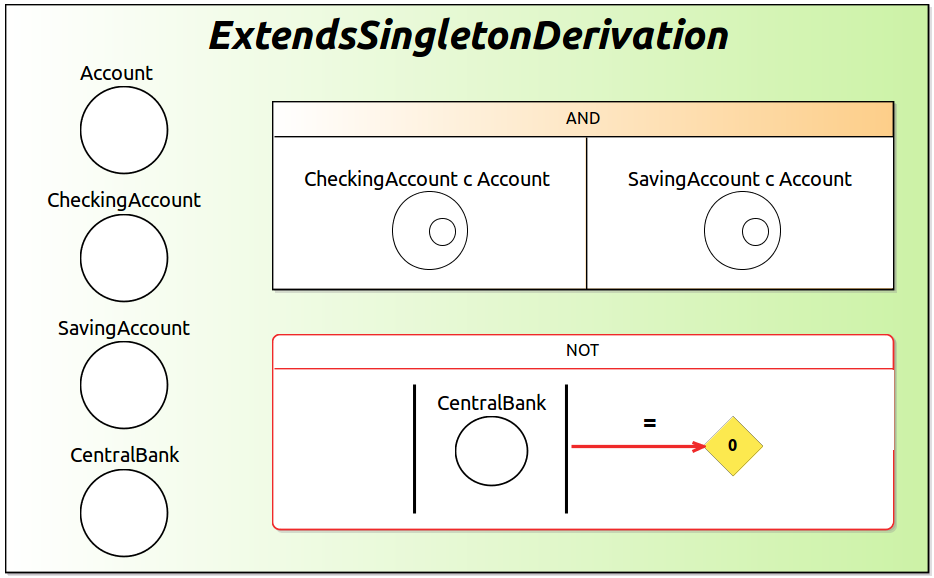}
    \caption{GIRL logical operation}
    \label{fig:GIRL_LogicalOperation}
\end{figure}

Figure~\ref{fig:GIRL_LogicalOperation} illustrates two logical operations that state that (i) as checking accounts as saving accounts are contained into an account; (ii) a central bank may not have zero instance.

\begin{figure}[]
    \centering
    \fbox{\includegraphics[width=0.45\textwidth]{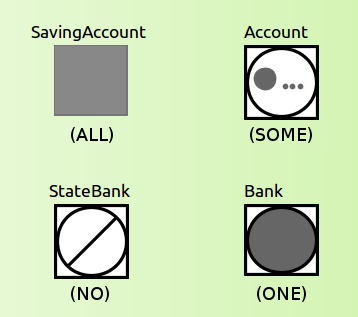}}
    \caption{GIRL quantifiers}
    \label{fig:GIRL_Quantifiers}
\end{figure}

\begin{figure}[]
    \centering
    \includegraphics[width=0.7\textwidth]{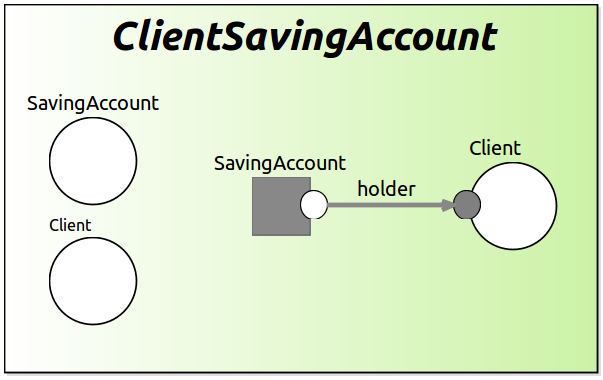}
    \caption{An example of GIRL quantification}
    \label{fig:GIRL_Quantification}
\end{figure}

\textbf{Quantification} - GIRL offers the existential and universal quantifiers as well as two derived quantifiers: the \textit{NO} and \textit{ONE} quantifiers. The former is the negation of the existential quantifier, whereas the latter restricts to exactly one instance that must obey the constraint. The notation for these four quantifiers is shown in Figure~\ref{fig:GIRL_Quantifiers}. As can be seen, their notation is similar to the multiplicity notation. Figure~\ref{fig:GIRL_Quantification} illustrates an example stating that all saving accounts must have only ONE holder client.

\begin{figure}[]
    \centering
    \fbox{\includegraphics[width=1\textwidth]{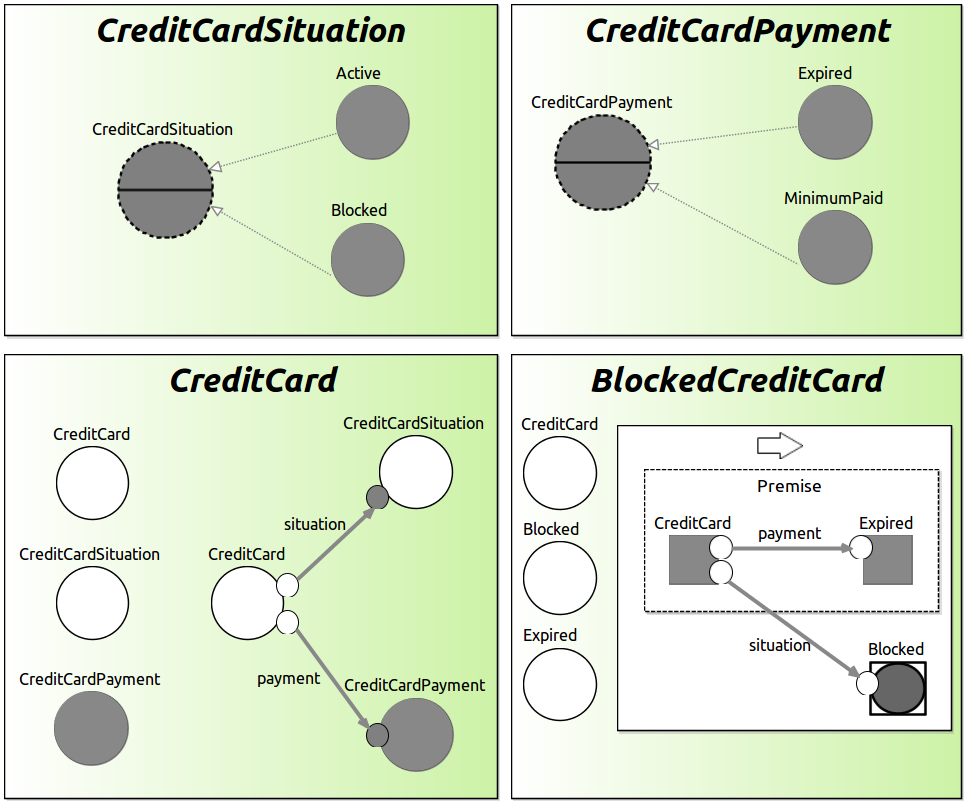}}
    \caption{GIRL implication}
    \label{fig:GIRL_Implication}
\end{figure}

\textbf{Implication} - The GIRL implication is applied on GIRL boolean expressions (premise and conclusion). The premise must have at least one quantifier. However, whenever a premise is a relationship, the conclusion must be necessarily also a relationship. The implication notation is a white rectangle with an arrow in the header, both with a black outline. The premise must be contained in a white rectangle with black dashed edges, whereas the conclusion comprises the remaining area outside of the premise rectangle. Figure~\ref{fig:GIRL_Implication} illustrates (in bottom right) the implication operator stating that if the credit card payment expires, the card must be blocked.

\begin{figure}[]
    \centering
    \includegraphics[width=0.7\textwidth]{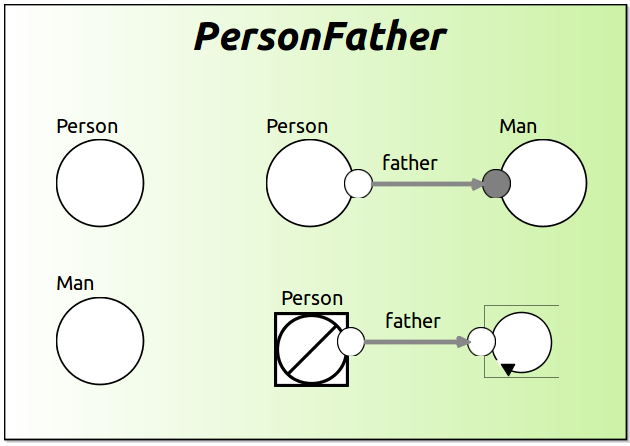}
    \caption{GIRL transitive closure operation}
    \label{fig:GIRL_TransitiveClosure}
\end{figure}

\textbf{Transitive Closure Operation} - GIRL provides the transitive closure on a relationship, as show in Figure~\ref{fig:GIRL_TransitiveClosure} whose invariant states that a person can not have himself among his male ancestors. You can note that this constraint is specified by stating that no person (\textit{NO} quantifier) is in the transitive closure of the relation \textit{father}.

\subsection{An Illustrative Example}
\label{example}
In order to better illustrate the GIRL language, we introduce in this section an example of a GIRL model built to meet the following four requirement constraint sets (RCS) specified on a graduate course domain:

\begin{itemize}
    \item RCS1: A graduate student may be classified as a  regular or as a special student; a professor may be classified as a permanent, as a collaborator or as a visiting professor;
    \item RCS2: A student must have exactly one concentration area; a professor must have one or more concentration areas;
    \item RCS3: A course may be classified either as a doctorate course or a masters course; a student can only be enrolled in one graduate course;
    \item RCS4: A student may have up to two advisor professors; if a student is enrolled in a doctorate course, so he/she must have one or more advisor professors.
\end{itemize}

\begin{figure}[]
    \centering
    \fbox{\includegraphics[width=0.9\textwidth]{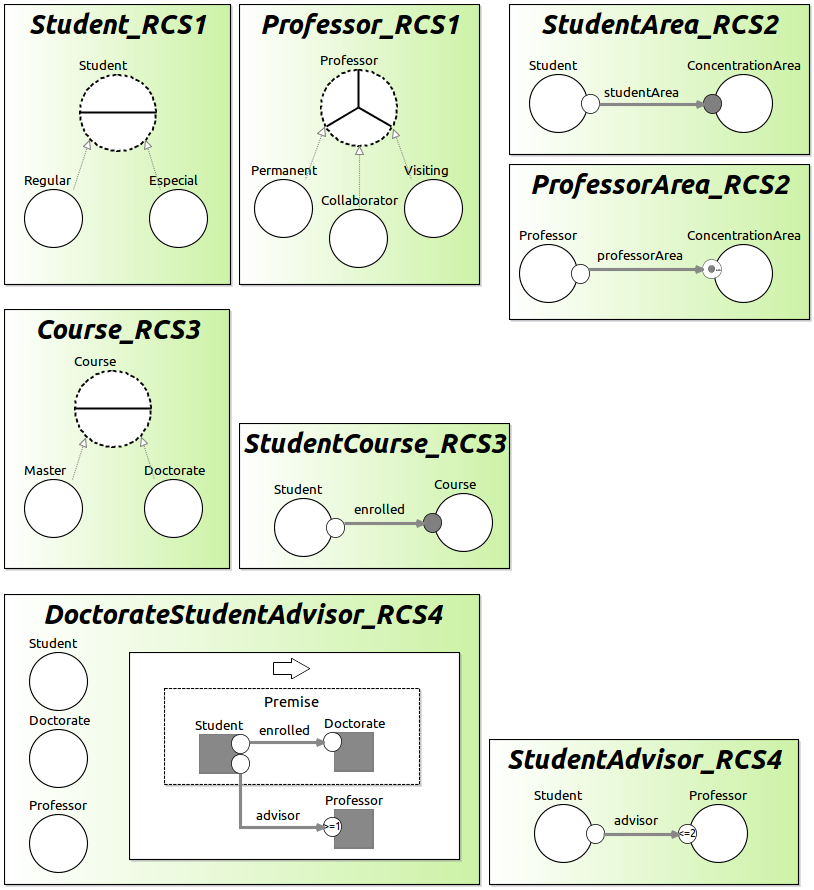}}
    \caption{An illustrative GIRL example}
    \label{fig:GIRL_IllustrativeExample}
\end{figure}

Figure~\ref{fig:GIRL_IllustrativeExample} shows a possible GIRL model built to specify the four aforementioned requirement constraint sets. On the left size, the \textit{Student} and \textit{Professor} entities are specified according to RCS1. As can be seen, they are specified as abstract entities that are extended according to RCS1 constraints. On the above middle part, two relationships are specified to model the RCS2 constraints, whereas on the bellow middle part an implication operator is applied to state the RCS4 constraints. Observe that the premise restricts the relationship only to doctorate students, and that \textit{ALL} quantification is applied on the three entities (\textit{Student}, \textit{Doctorate} and \textit{Professor}). Finally, on the above right side five entities specify RCS3 constraints, whereas on the bellow right side one relationships still states the RCS4 constraints.

\section{Mapping Semantics}
\label{map}
By means of a mapping semantics, we define GIRL models as a module of an Alloy specification. 
This mapping requires processing on some structures to avoid redundancy in the generated Alloy model. 
The mapping is implemented using MDA transformations (Section~\ref{tool}). 
In this section, we introduce this mapping by illustrating  GIRL constructs (on the left-hand side) and its correspondent Alloy code (on the right-hand side).

\subsection{Invariant $\rightarrow$ Fact}
A GIRL invariant is mapped to an Alloy fact comprising the expressions that are recursively mapped.

\subsection{Entity $\rightarrow$ Signature}
\begin{figure}[]
    \centering
    \includegraphics[width=0.55\textwidth]{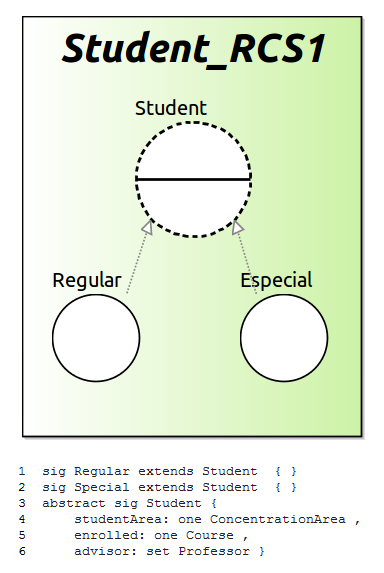}
    \caption{GIRL Entity to Alloy signature mapping}
    \label{fig:GIRL_EntitySignatureMapping}
\end{figure}

A GIRL entity is mapped to an Alloy signature. In case of being an abstract or a singleton entity, the generated Alloy signature must contain the \textit{abstract} or \textit{singleton} attribute, respectively. Figure~\ref{fig:GIRL_EntitySignatureMapping} illustrates this mapping. 
On the right side, the signature in Lines 1-2 states that \textit{Regular} and \textit{Special} signatures extend the \textit{Student} abstract signature (Lines 3-6). 
For each entity on the left-hand side, an Alloy signature is provided on the right-hand side.

\subsection{Relationship $\rightarrow$ Signature}

For each GIRL relationship, Alloy attributes are created, where:
\begin{itemize}
    \item the source entity from the GIRL relationship is the owner of the Alloy relation, and the Alloy signature field name is the same of the GIRL relationship target element, given the relationship does not have quantifiers or one of the following quantifiers:  \textit{one}, \textit{lone}, \textit{some} and \textit{set}, applied on the target element. For instance, in Figure~\ref{fig:GIRL_EntitySignatureMapping} , the \textit{studentArea} is generated into the \textit{Student} signature. 
    However, if the target element is a relational expression, a quantifier expression is generated for the invariant;
    \item The type of the generated field is the Alloy signature generated to the target element from the GIRL relationship. In case of the relationship being also applied to an abstract entity extended by the target entity, the abstract entity must be the type of the field. For instance, in Figure~\ref{fig:GIRL_EntitySignatureMapping}, the type of \textit{studentArea} attribute is \textit{ConcentrationArea}. 
    On the other hand, if the target element is a set operation, the type of the field is an Alloy set operation containing the signatures generated to the source and target elements of the relationship;
    \item the GIRL relationship source element generates an Alloy expression with the multiplicities \textit{one}, \textit{lone} or \textit{some} on the inverse relation, as is shown in Figure~\ref{fig:GIRL_EntitySignatureMapping}, where the \textit{studentArea} is generated into the \textit{Student} signature with the \textit{one} quantifier. However, whenever a multiplicity is constrained utilizing a relational operator (\textit{e.g.} $\leq$ 2), a quantifier expression is generated, comparing the cardinality in the inverse relation. Both expressions are associated with the fact generated to the invariant containing the relationship.
\end{itemize}

\subsection{Remaining Operations $\rightarrow$ Alloy Expressions}

\begin{figure}[]
    \centering
    \includegraphics[width=0.65\textwidth]{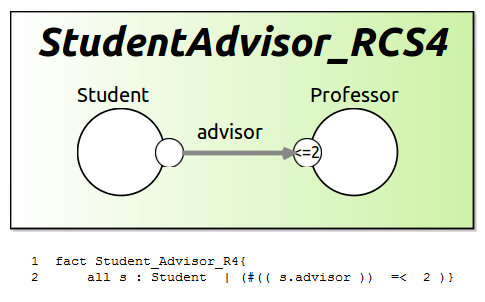}
    \caption{Relational operation to Alloy fact mapping}
    \label{fig:GIRL_RelationalOperationMapping}
\end{figure}

A GIRL relational operation is mapped to an Alloy fact comprising a comparison expression, where the relational operator is directly mapped between both. Figure~\ref{fig:GIRL_RelationalOperationMapping} illustrates this mapping using the constraint stating that a student may have up to two advisor professors. 
The previously described GIRL model is shown on the left side, whereas the Alloy fact to which the former is mapped to is shown on the right side. 
It is specified that for all variables\footnote{The Alloy analyzer API generates anonymous variables whose long unique ids are simplified here.} that instantiate a \textit{Student}, there is a constraint stating that its \textit{advisor} attribute cardinality must be lesser or equal than 2. 

A GIRL implication is mapped to an Alloy fact comprising an implies expression. 
In the case of an existing relationship in the implication, the argument must be an Alloy quantification expression comprising the comparison expression with the operator \textit{in} as in the premise as in the conclusion of the implication. 
Otherwise, the arguments must be an Alloy implication whose premise and conclusion are Alloy expressions associated with containment, relational, or logical operation according to the aforementioned mappings.

\begin{figure}[]
    \centering
    \includegraphics[width=1\textwidth]{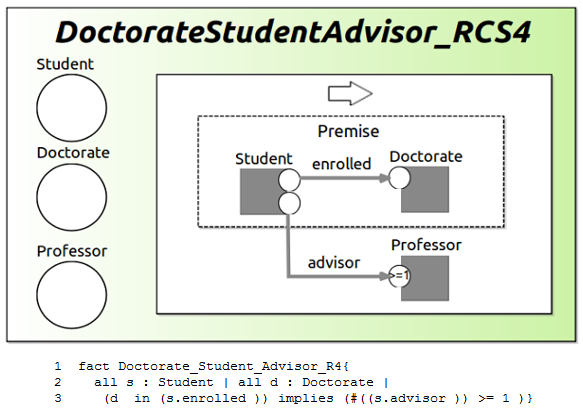}
    \caption{Implication to Alloy fact mapping}
    \label{fig:GIRL_ImplicationOperationMapping}
\end{figure}

Figure~\ref{fig:GIRL_ImplicationOperationMapping} shows the Alloy code (right-hand side) to which is mapped the GIRL model excerpt (left-hand side) that states if a student is enrolled in a doctorate course, so he/she must have one or more advisor professors. An Alloy \textit{all} quantification is generated, comprising the relationship with the operator \textit{in} for the premise and the relational operation for the conclusion.

A GIRL quantification is mapped to an Alloy quantification where the quantifier type is directly mapped between both because Alloy provides the \textit{all}, \textit{one}, \textit{some} and \textit{no} quantifiers. Figure~\ref{fig:GIRL_ImplicationOperationMapping} above illustrates this mapping, where both GIRL \textit{all} quantifiers are directly mapped to the \textit{all} Alloy quantifiers.
A GIRL logical operation is mapped to (i) an Alloy expression with the conjunction or the disjunction operator; or (ii) an Alloy unary expression with the negation operator.

\section{Tool Support}
\label{tool}

In this work, we provide a prototypical IDE (Integrated Development Environment) to allow users  adopt GIRL as visual language to specify requirement invariants, taking benefit from its expressiveness as from the Allow solver pursued as target language to detect inconsistences in the GIRL model. 
Any GIRL model may be created by means of a model graphical editor implemented with the Sirius~\cite{sirius:2019} along with the EMF Framework~\cite{emf:2019}. The whole GIRL visual elements are provided by the editor, implemented as an Eclipse plugin. 

In order to analyze GIRL models, detecting their potential inconsistencies, 
we implemented the mapping rules presented in Section~\ref{map} as a set of QVT transformations from GIRL to Alloy.
These transformations receive GIRL EMF models as input and generate Alloy EMF models as output, both according to the GIRL and Alloy meta-models, respectively.
The Alloy EMF is thus transformed into a textual Alloy specification by means of Acceleo textual transformations. Therefore, the Alloy code feeds the Alloy SAT-Solver by means of the available API~\cite{AlloyAPI:2019}. 
The prototype analyzes a GIRL model by means of the Alloy SAT Solver. 
As a result, the model is showed as satisfiable or not. In case of being satisfiable, entity and relationship instances are exhibited. Otherwise, counterexamples are shown.



\begin{figure}[]
    \centering
    \includegraphics[width=0.9\textwidth]{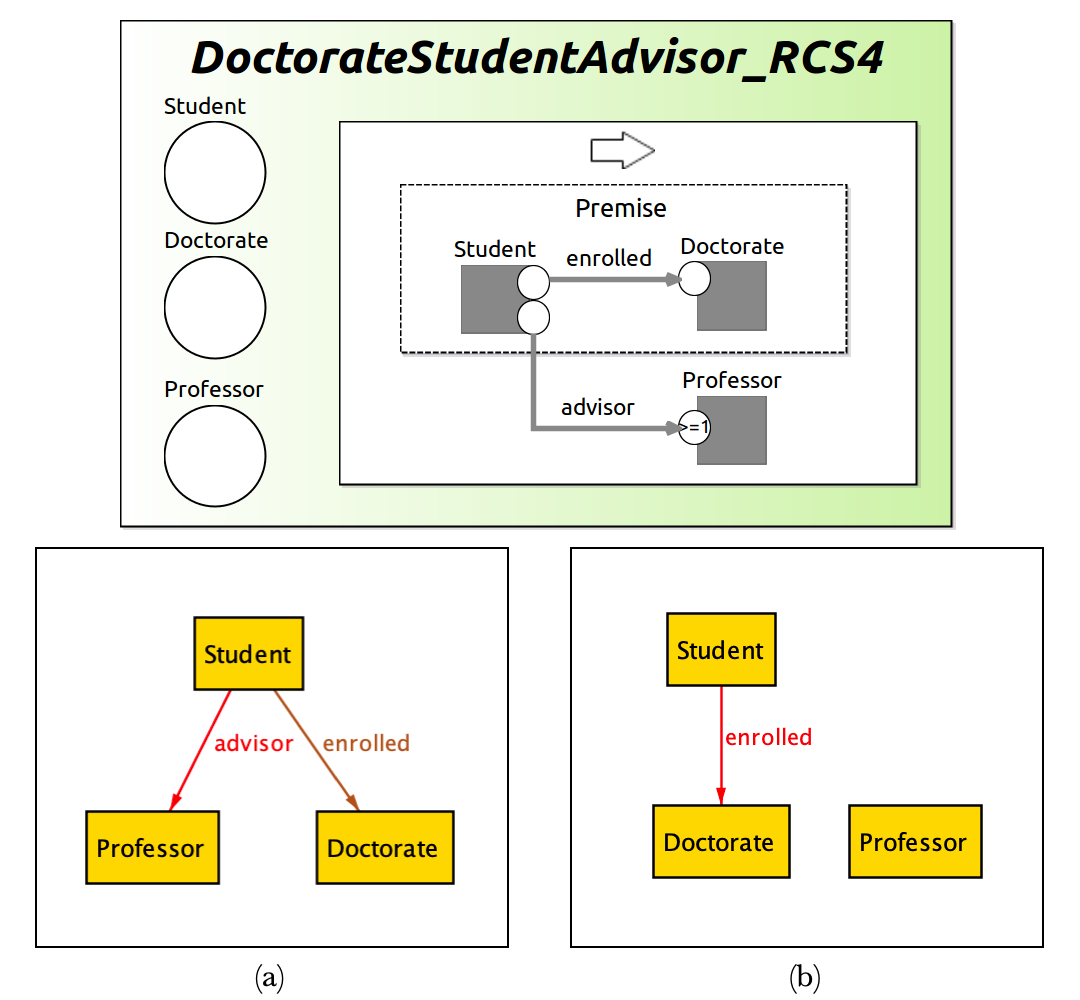}
    \caption{Automatic analysis example.}
    \label{fig:tool-example}
\end{figure}

As an example, Figure~\ref{fig:tool-example} shows two possible analysis cases for the constraint in RCS4 (Section~\ref{example}), ``if a student is enrolled in a doctorate course, so he/she must have one or more advisor professors". Figure~\ref{fig:tool-example}(a) is one resulting instance from the example, which is provided by the IDE without any test input. On the other hand, assuming the analyst did not specify that constraint correctly, the instance in Figure~\ref{fig:tool-example}(b) is shown, signaling that an unwanted scenario is being allowed (a student enrolled in the doctorate program but not linked with any advisors); the visualization of this kind of modeling problems is valuable for detecting requirement inconsistencies early in the process.

\section{Study Methodology}
\label{eval}

We evaluate GIRL through an empirical study with requirement analysts, working for a government software development company in Brazil. During their tasks, we assess GIRL's effectiveness and usability. 

\subsection{Definition}
\label{rqs}

The main aim of this study is to evaluate GIRL in representing and automatically checking structural invariants in software requirements, with the intention of characterizing the application of formal analysis concealed behind a graphical DSL, with respect to the effectiveness of its use in terms of specification quality and degree of difficulty in language learning. The study was carried out from the point of view of requirements analysts, in the context of the elicitation, analysis and specification of structural invariants for information systems.

The following research questions guide our discussion:

\textit{RQ1: How hard is it to learn and use GIRL?} We intend to characterize how natural GIRL is when used by requirement analysts.

\textit{RQ2: How are accuracy and coverage of the requirements represented with GIRL?} We discuss whether the acquired knowledge has been appropriately applied, through the resulting requirements.

\textit{RQ3: How effective is GIRL in the automatic verification of requirements?} We evaluate how requirement analysts employ the automatic detection of inconsistencies provided by GIRL.


\subsection{Context}
As participants, we selected software professionals who had performed requirement elicitation activities, regardless of the used development process. Those professionals were recruited from a Brazilian government organization that develops software for Brazil's public universities. After invitations sent by e-mail, 12 professionals accepted to participate voluntarily. We applied a pilot study to two of them at first, then 10 participated in the final study.
Table~\ref{tab:demog} presents some characteristics per participant as collected with an invitation questionnaire.
We found that most have considerable experience in requirements (from 6 to 10 years) and self-assessed intermediate to advanced knowledge in logic.
Also, the questionnaire asked for session schedule suggestions. In agreement with the participants, we scheduled one participant per session.

\begin{table}[]
\centering
\caption{Participant demographics.}
\label{tab:demog}
\resizebox{0.8\textwidth}{!}{%
\begin{tabular}{|l|l|l|l|}
\hline
\textbf{Part.} & \textbf{\begin{tabular}[c]{@{}l@{}}Experience with \\ requir. analysis\end{tabular}} & \textbf{\begin{tabular}[c]{@{}l@{}}Knowledge about \\ Math. Logic\end{tabular}} & \textbf{\begin{tabular}[c]{@{}l@{}}Degree \\ Level\end{tabular}} \\ \hline
P01            & None                                                                                 & Advanced                                                                        & Bachelor                                                         \\
P02            & 6 to 10 years                                                                        & Intermediate                                                                    & Bachelor                                                         \\
P03            & 6 to 10 years                                                                        & Advanced                                                                        & Bachelor                                                         \\
P04            & 6 to 10 years                                                                        & Intermediate                                                                    & MSc                                                              \\
P05            & 6 to 10 years                                                                        & Intermediate                                                                    & MSc                                                              \\
P06            & 3 to 5 years                                                                         & Intermediate                                                                    & MSc                                                              \\
P07            & 6 to 10 years                                                                        & Intermediate                                                                    & Bachelor                                                         \\
P08            & 1 to 2 years                                                                         & Advanced                                                                        & MSc                                                              \\
P09            & 6 to 10 years                                                                        & Intermediate                                                                    & MSc                                                              \\
P10            & 6 to 10 years                                                                        & Intermediate                                                                    & MSc                                                              \\ \hline
\end{tabular}%
}
\end{table}

To train all participants in the GIRL environment, we employed invariant examples from a typical banking system, covering the primary constructs (relationship, logical operation and quantification). As an instrument, we used video training, followed up by a set of questions using the banking system.

For the main study, the participants were asked to model four sets of invariants (RCS1 to RCS4 from Section~\ref{tool}), taken from a real system developed by the organization; before selection, we ensured no participants had worked with that system before.

\subsection{Experimental Procedure}

Following the classification proposed by Stol e Fitzgerald~\cite{StolFitzgerald}, our study follows the \emph{Judgment Task} strategy, in which participants are asked to judge and measure behavior or discuss topics of interest. The sample is systematic, although non-representative.
The study took place at two moments: first, we applied a pilot study to validate instruments and planning. Next, the main study was composed of individual sessions for each participant. 
Each session had three stages: (i) setting and training; (ii) use of the GIRL language to represent requirements, and (iii) a post-study evaluation of GIRL and the entire experience.
At any moment, the participant had free access to the material available on the language, but also could present his/her questions to an experienced researcher, who was always present.

The main task was carried out by modeling in GIRL four sets of structural invariants. While RCS1 required modeling \emph{Entity} and \emph{EntityAbstract} constructs, RCS2  induced participants to use \emph{Relationship} and multiplicities. RCS3 required the use of logical expressions,  then RCS4 demanded \emph{Quantification} and \emph{Implication}.
Participants were not asked to follow a predetermined order for working on those sets. 
In each session, the researcher stimulated the participant to share their thoughts using the Think-Aloud protocol~\cite{Think-aloud}. Besides recording the session, the researcher took notes on (i) participant questions and comments about GIRL, (ii) participant's body expression, especially fatigue or boredom, (iii) verbalization of participant's reasoning when modeling invariants.

We maintained an identical physical environment for all participants. The room presented appropriate lighting and a chart with explanatory reminders about GIRL; also, internet-enabled computer, the GIRL environment (Eclipse plugin) open on an individual, empty workspace, GIRL tutorial, both in print and PDF and other supplies, such as pen and paper.
The researcher explained the procedure for each step of the study, recording the start and end times of each session.
After finishing his/her version of the model, the participant was handed a printed suggested solution to the four sets of invariants, next invited to compare the solutions and comment on the result.

We employed a mixed approach to answering the research questions. Quantitative data come up from direct measures and the post-study questionnaire, namely number of inquiries, time to represent requirements, and degree of usefulness for the tool. 
Likewise, the qualitative data from the think-aloud comments and the post-study questionnaire was analyzed by two of the authors, independently, and categorized in the following topics:
\begin{itemize}
    \item GIRL;
    \item Domain;
    \item Analysis feedback;
    \item IDE.
\end{itemize}
 
\section{Results and Discussion}
\label{results}

We present and discuss results from the study in terms of the research questions established in Section~\ref{rqs}.

\subsection{RQ1: How hard is it to learn and use GIRL?} 

At first, participants assessed their preparatory training as \emph{high} -- from 7 to 10 -- for all structures in GIRL; the questions assessed their learning of entities, cardinality, relationships, containment, set operations, logical operations, invariants, quantification, and implication. 
After modeling the four sets of invariants, the participants were asked to grade GIRL's usefulness in the task. Table~\ref{tab:quantitative} shows the quantitative data we collected during the study.

\begin{table}[]
\centering
\caption{Quantitative data.}
\label{tab:quantitative}
\resizebox{0.6\textwidth}{!}{%
\begin{tabular}{|l|r|r|r|}
\hline
\textbf{Part.} & \multicolumn{1}{l|}{\textbf{\begin{tabular}[c]{@{}l@{}}Elapsed\\ Time\\ (approx. \\ in min)\end{tabular}}} & \multicolumn{1}{l|}{\textbf{\begin{tabular}[c]{@{}l@{}}Inquiries\\ during\\ Task\end{tabular}}} & \multicolumn{1}{l|}{\textbf{\begin{tabular}[c]{@{}l@{}}After-study\\ overall\\ GIRL\\ evaluation\end{tabular}}} \\ \hline
P01            & 49                                                                                                         & 11                                                                                              & 9                                                                                                               \\ \hline
P02            & 31                                                                                                         & 3                                                                                               & 9                                                                                                               \\ \hline
P03            & 34                                                                                                         & 12                                                                                              & 10                                                                                                              \\ \hline
P04            & 57                                                                                                         & 4                                                                                               & 8                                                                                                               \\ \hline
P05            & 58                                                                                                         & 12                                                                                              & 10                                                                                                              \\ \hline
P06            & 52                                                                                                         & 10                                                                                              & 9                                                                                                               \\ \hline
P07            & 37                                                                                                         & 8                                                                                               & 9                                                                                                               \\ \hline
P08            & 39                                                                                                         & 4                                                                                               & 9                                                                                                               \\ \hline
P09            & 71                                                                                                         & 11                                                                                              & 10                                                                                                              \\ \hline
P10            & 37                                                                                                         & 9                                                                                               & 9                                                                                                               \\ \hline
\end{tabular}%
}
\end{table}

Except for one participant, all took less than an hour to complete.
By the end of the task, although the level of complexity was considered low for simpler structures (entity, relationship and invariant), participants reported problems in more complex structures (quantification, implication, cardinality, and containment).  
Their inquiries focused on concepts not commonly used in requirement elicitation tasks, such as \emph{``Implication isn't intuitive, we must be aware of the order"}.
Despite the appeal of a graphical modeling language, users still seemed to struggle with the semantics of logical implication and quantification.
The challenge is to come up with an intuitive graphical representation of those constructs in invariants.

Nevertheless, most participants built models with negligible observer intervention. 
In fact, they registered a high degree of language usefulness (Table~\ref{tab:quantitative}). 
Table~\ref{tab:girl-comments} lists the most relevant topics extracted from the qualitative data.

\begin{table}[]
\centering
\caption{Qualitative data about the GIRL language.}
\label{tab:girl-comments}
\resizebox{0.9\textwidth}{!}{%
\begin{tabular}{|l|l|}
\hline
\textbf{Part.} & \textbf{Main excerpts and inquires about GIRL}                                                                                                                                                                                                \\ \hline
P01            & ``It's hard to know when to apply quantifiers and invariants"                                                                                                                                                                                  \\ \hline
P02            & \begin{tabular}[c]{@{}l@{}}``Cardinality, quantification and invariants, I may \\  have got those wrong"\\``I may have low experience with theory"\end{tabular}                                                                                \\ \hline
P03            & \begin{tabular}[c]{@{}l@{}}``some structures could be easier to draw (...) for implication, \\ we should select the premise first, draw it on one area in the canvas, \\ then the conclusion in a second area.\end{tabular}                    \\ \hline
P05            & \begin{tabular}[c]{@{}l@{}}``Found it very intuitive, the structures, and the graphical \\  representation of the concepts."\\ ``It was hard to remember how to quantify inside an implication"\end{tabular}                                    \\ \hline
P06            & \begin{tabular}[c]{@{}l@{}}``Invariants, relationships and multiplicities were pretty \\ straightforward to use"\\ ``Using implications and quantifications was the most complex"\end{tabular}                                                  \\ \hline
P07            & \begin{tabular}[c]{@{}l@{}}``The easiest structures to use were entity and relationship, since they \\  are similar in other modeling languages."\\ ``Implication isn't intuitive, we must be aware of the order."\end{tabular}                 \\ \hline
P08            & \begin{tabular}[c]{@{}l@{}}``Structures like entity, relationship, and implication were easier to \\ understand"\\ ``I found it hard to tell containment from relationship"\end{tabular}                                                        \\ \hline
P09            & \begin{tabular}[c]{@{}l@{}}``Regarding structures, they were easy to learn and model requirements. \\ Also, it is good to have a complete set of components in the language, \\ to model complex requirements easily and clearly"\end{tabular} \\ \hline
P10            & \begin{tabular}[c]{@{}l@{}}``Like the language's expressiveness."\\ ``Overloading of constructs raises inquires about \\  their use (e.g. cardinality)"\end{tabular}                                                                            \\ \hline
\end{tabular}%
}
\end{table}

\subsection{RQ2: How are accuracy and coverage of the requirements represented with GIRL?} 

Table~\ref{tab:req-comments} shows the main comments related to the modeled requirements.
Most comments were made after the participants finished their tasks and were able to check their resulting model against a suggested answer provided by the mediator.

\begin{table}[]
\centering
\caption{Qualitative data about modeling the study's requirements.}
\label{tab:req-comments}
\resizebox{0.9\textwidth}{!}{%
\begin{tabular}{@{}|l|l|@{}}
\toprule
\textbf{Part.} & \textbf{Main excerpts and inquires about requirement modeling}                                                                                                                                                                                                                                                                    \\ \midrule
P01            & \begin{tabular}[c]{@{}l@{}}``I believe I did not understand quantifiers together with implications"\\ ``I got `having one or more' wrong"\end{tabular}                                                                                                                                                                     \\ \midrule
P02            & ``I have modeled in a different way, but I guess it is correct"                                                                                                                                                                                                                                                           \\ \midrule
P03            & \begin{tabular}[c]{@{}l@{}}``My proposal constrained the course possibilities a little more"\\ ``I could have made Student and Professor in a different way, \\ a more reusable form"\end{tabular}                                                                                                                         \\ \midrule
P04            & \begin{tabular}[c]{@{}l@{}}``I considered two additional relationships course\_type and \\ professor\_type, not in the suggested answer"\\ ``I did not model left multiplicity correctly"\end{tabular}                                                                                                                     \\ \midrule
P05            & ``I didn't consider multiplicity for the entity in R4"                                                                                                                                                                                                                                                                   \\ \midrule
P06            & \begin{tabular}[c]{@{}l@{}}``I thought of using singleton to model R4, but it made it \\ harder to model"\end{tabular}                                                                                                                                                                                                 \\ \midrule
P07            & \begin{tabular}[c]{@{}l@{}}``For R3, I modeled the relationship between course and \\ student in a different, less intuitively than the solution suggested;"\\ ``In R4, for the relationship between phd-student and advisor, the \\ multiplicity is neglectable, since the quantification is already there."\end{tabular} \\ \midrule
P09            & \begin{tabular}[c]{@{}l@{}}``I used a more complex representation, using a few more \\ language constructs, for the same constraints"\end{tabular}                                                                                                                                                                        \\ \midrule
P10            & \begin{tabular}[c]{@{}l@{}}``I used less entities in my model; I believe this will make \\ analysis harder, due to the different granularity in \\ the representations"\end{tabular}                                                                                                                                     \\ \bottomrule
\end{tabular}%
}
\end{table}

Almost all participants represented the eight invariants correctly.
Only one participant (P08) defined one invariant incorrectly, by misusing a containment operator; however, right away the participant fixed it to the expected relationship.
Although having problems in understanding how to use implication correctly, they all realized the correct need for implications in the invariants.
However, it was more difficult for them to realize the need for quantifiers.
The reason for this result may be due to the lack of association between the graphical representation of quantification with the concept, or even the lack of insight on the concept itself. 

P02, P03, P06 and P07 are examples of those who understood the language allows for several possible models for the same constraint. Their willingness to accept this difference probably came from the automatic analysis. The alternative forms may, however, make it hard to visualize the solution shown by the automatic analysis, as P10 comments suggest.

In a few cases, participants made wrong assumptions about the semantics of GIRL constructs.
P01 built an under-constrained model, by not enforcing a one-or-more relationship, which was not detected by the analysis tool.
Similarly, P03 produced an over-constrained model.
We believe that, despite the automatic visualization of instances from the model, users may need additional training in looking for those differences; this skill does not seem to be straightforward in software analysts, as observed in \cite{readability-formal,human-factors-formal}.
Furthermore, the commentary by P04 shows an interesting aspect about modeling languages: beginners tend to use more features than needed. The skill to keep models minimalist requires more experienced modelers, independently of the formal infrastructure used by the language.

\subsection{RQ3: How effective is GIRL in the automatic verification of requirements?} 
According to Table~\ref{tab:quantitative}, the after-study assessment was high for all participants (8-10), in terms of how GIRL and its infrastructure was useful for understanding the requirements. 
In Table~\ref{tab:analysis-comments}, we highlight the fragments related to the automatic analysis, reported by participants.

\begin{table}[]
\centering
\caption{Fragments reported about the automatic analysis}
\label{tab:analysis-comments}
\resizebox{0.8\textwidth}{!}{%
\begin{tabular}{|l|l|}
\hline
\textbf{Part.} & \textbf{Main excerpts about automatic analysis}                                                                                                                                                     \\
\hline
P01            & \begin{tabular}[c]{@{}l@{}}``With the analysis, I saw the `course' relation as duplicate"\\ ``Oh...a master student must have an advisor"\\ ``Extremely useful, especially if executed incrementally. "\end{tabular}        \\ \hline
P02            & \begin{tabular}[c]{@{}l@{}}``Saw that I changed `has' relation with `enrolled'"\\ ``Good to have discovered inconsistencies in modeling"\end{tabular}                                                                      \\ \hline
P03            & ``Verification was important to see a Professor in two areas"                                                                                                                                                             \\ \hline
P04            & \begin{tabular}[c]{@{}l@{}}``Found out with the verification two relations with the \\ same name"\\ ``Analysis found out that I had modeled a student \\ having exactly one course, which is too restrictive"\end{tabular} \\ \hline
P05            & \begin{tabular}[c]{@{}l@{}}``One student is enrolled in one or more courses"\\ ``A course cannot have zero students?"\\ ``Found out that Professor cannot be a singleton"\end{tabular}                                      \\ \hline
P06            & \begin{tabular}[c]{@{}l@{}}``One modeled relation was unnamed"\\ ``I saw a student that is enrolled in nothing"\\ ``Now I see Master student with no advisor"\end{tabular}                                                  \\ \hline
P07            & \begin{tabular}[c]{@{}l@{}}``I see I did not allow more Courses for one Student"\\ ``That would be a good way to analyze my use cases"\\ ``It is like debugging"\end{tabular}                                               \\ \hline
P08            & ``Analysis was useful for evaluating the model"                                                                   \\ \hline
P09            & \begin{tabular}[c]{@{}l@{}}``It is different, when I program, after I write \\ all the code I test"\\ ``Does it say how many solutions there are?"\end{tabular}                                                            \\ \hline
P10            & \begin{tabular}[c]{@{}l@{}}``I'd like a better way to navigate through the solutions"\\ ``Got an inconsistency from the analysis; I \\ overconstrained a relation to `Area'"\end{tabular}                                  \\ \hline
\end{tabular}%
}
\end{table}

By analyzing the qualitative data, all participants made at least one positive remark about the automatic visualization of results, as provided by the Alloy's formal infrastructure. As the requirements were seen in action, participants were able to get feedback that made them quickly fix inconsistencies; these events are explicitly stated in fragments by P01, P03, P04 and P10. Furthermore, P01 and P07 reported the benefit of \textit{incremental modeling}, in which analysis is automatic with no need for parameter input.

We noticed high expectation to see the results being shown, in general. To most of them, it was novel to use any tool for analyzing requirements, which is perceived as an issue in the formal methods community~\cite{jackson2000software,formal-practice}.
Two participants (P09 and P10) suggested more control over all instances of a model; this was not offered in the IDE due to a limitation of Alloy's analysis, which finds an instance for each simulation execution.


\subsection{Threats to Validity}

We discuss the limitations of this study by using the categories listed by Wohlin et al.~\cite{Wohlin-book}.

\noindent\textbf{Conclusion validity.} Our small sample (10 participants) may certainly affect conclusions. Nevertheless, the study has been designed to ensure replicability for larger samples, and we prioritized qualitative data. Regarding the questionnaires, we tried to establish clarity by a pilot study with two analysts; also, we split the questions in shorter modules, to minimize fatigue.
The mediating researcher tried to take notes from the think-aloud protocol as soon as they were expressed by participants, and asked for clarification in case of unclear comments.

\noindent\textbf{Internal validity.} One kind of bias we tried to avoid is the likely interaction among participants; we scheduled individual sessions, asking participants to keep away of exchanging information with other participants through the execution of all sessions. Furthermore, we tried to avoid participants to disengage from the study by offering a flexible schedule for the individual sessions.

\noindent\textbf{External validity.} The selected participants are software developers with varying experience levels with requirement analysis; although the conclusions cannot be generalized -- as we could not use a representative sample -- we assume the study is exploratory, with emphasis on qualitative data. Also, despite we used a simulated environment, we applied requirements from a real system.

\section{Related Work}
\label{related}
In a systematic review, Gonzalez et al.~\cite{cabot14} evaluate approaches to verify models. Among 18 studied approaches, only one used Object Oriented Data Base Schema. The remaining ones used UML class diagrams. Most of them, also adopted OCL to specify constraints on the UML model. In some of these approaches \cite{uml2alloy,EMFtoCSP,cdtoalloy,uml-ocl18,uml-b}, models are automatically verified by means of diverse approaches, as SAT (Boolean Satisfiability), CSP (Constraint Satisfaction Problem) and SMT (Satisfiability Modulo Theories). However, eight (8) of them do not provide feedback to the users.
Several papers have proposed different mappings from UML with OCL to a formal language in order to allow verification of its soundness, such as UML2Alloy \cite{uml2alloy}, CD2Alloy \cite{cdtoalloy} , EMFToCSP \cite{EMFtoCSP}, CD2Formula \cite{uml-ocl18} and UML-B \cite{uml-b}. However, it is important to highlight that OCL is a textual language, differently from GIRL.

On the other hand, three (3) studies~\cite{constraint-diags,constraint-trees,visual-uc} have proposed graphical languages to model invariants.
Kholkar et al.~\cite{visual-uc} propose a visual language named Business Rule Diagram to express invariants. This diagram allows to model business rules associated to objects. 
However, it does not provide implication and quantification operations, and logical operations are provided with textual syntax. The verification is provided by algorithms based on Binary Decision Diagram. 
Kent and Howse~\cite{constraint-trees} extend the Constraint Diagrams to UML previously proposed in \cite{constraint-diags}. Their Constraint Trees \cite{constraint-trees} incorporate OCL in its visual notation. It is also based on Venn diagrams. However, it does not provide implication operation. GIRL and Constraint Trees provide similar concepts, such as singleton sets and cardinalities. However, there are differences in the way they visually provide them. For instance, Constraint Trees require that singleton sets are inside a set. Additionally, Constraint Trees demand these structures to be represented in a class diagram. Currently, there is no tool support to manipulate Constraint Trees as well as to automatically check the models.

Moody suggests~\cite{physics} the use of nine principles called Physics of Notation (PoN) that establish cognitively effective visual notations based on disciplines from other fields of knowledge such as communication, semiotics, graphic design, visual perception, and cognition. These principles inspired the visual notation of GIRL as well as they were used in some way by other studies, as in~\cite{linden17}.

\section{Conclusions}
\label{conclusions}
In this work, we introduced GIRL, a Domain-Specific Language to specify structural constraints in software requirements. 
GIRL's design goal is to establish a simple visual language, based on set notation, in which requirement constraints can be subject to automatic analysis. 
This analysis is provided by the Alloy Analyzer infrastructure; we provide a mapping semantics for GIRL constructs using the Alloy language, which was implemented in a prototypical IDE. This mapping allows for the detection of inconsistencies early during the requirement specifications, offering the benefits of formal analysis, but hiding a more mathematical-based formal language from analysts.

We evaluated GIRL and its automatic analysis by a mixed empirical study with ten voluntary software developers within a government-based software company in Brazil. By applying a judgment task~\cite{StolFitzgerald} methodology, participants were asked to model, in GIRL, four sets of requirement constraints from a real system, expressing their impressions and reactions by the think-aloud protocol. 
As result, participants did not report problems in using entities and relationships, although implication and quantification, even in a graphical form, were seen as complex. Nevertheless, the automatic analysis allowed them to detect inconsistencies, and 9 out 10 specified all constraints correctly.

As future work, 
we aim to extend GIRL in order to allow checking also the software behaviour. 
In addition, the visual notation can be improved, mainly concerning those constructs indicated in the empirical study as the more complex to understand. 
Also, we want to improve the prototypical IDE with features that were requested by participants during the study, such as auto-complete to entity and relationships, and automatic rearranging of the content of composite structures. 
Finally, a more complete evaluation is required mainly to cover all GIRL elements and apply them in real large-scale software projects.

\bibliographystyle{model1-num-names}

\end{document}